\documentclass[10pt]{article}
\usepackage[OE]{express}

\usepackage{graphicx}
\usepackage{dcolumn}
\usepackage{bm}
\usepackage{float}
\usepackage{amsmath}
\usepackage{braket}

\newcommand{\un}[1]{\mathrm{\:#1}}

\begin{document}

\title{Fast Generation and Detection of Spatial Modes of Light using an Acousto-Optic Modulator}
\author{Boris Braverman\authormark{1,*}, 
	Alexander Skerjanc\authormark{1}, 
	Nicholas Sullivan\authormark{1}, 
	Robert W. Boyd\authormark{1,2}}

\address{\authormark{1}Department of Physics and Max Planck Centre for Extreme and Quantum Photonics,	University of Ottawa, 25 Templeton Street, Ottawa, ON, K1N 6N5, Canada}
\address{\authormark{2}Department of Physics and Astronomy, University of Rochester, 275 Hutchison Road, Rochester, NY, 14627, USA}
\email{\authormark{*} bbraverm@uottawa.ca}

\date{\today}
\begin{abstract*}
	Spatial modes of light provide a high-dimensional space that can be used to encode both classical and quantum information. Current approaches for dynamically generating and measuring these modes are slow, due to the need to reconfigure a high-resolution phase mask such as a spatial light modulator or digital micromirror device. 
	The process of updating the spatial mode of light can be greatly accelerated by multiplexing a set of static phase masks with  a fast, image-preserving optical switch, such as an acousto-optic modulator (AOM).
	We experimentally realize this approach, using a double-pass AOM to generate one of five orbital angular momentum states with a switching rate of up to $500 \un{kHz}$.
	We then apply this system to perform fast quantum state tomography of spatial modes of light in a 2-dimensional Hilbert space, by projecting the unknown state onto six spatial modes comprising three mutually unbiased bases. We are able to reconstruct arbitrary states in under $1 \un{ms}$ with an average fidelity of 96.9\%.
\end{abstract*}


\section{Introduction}

The application of spatial of modes of light to classical and quantum communication is an area of active research, because of the potential for increased energy efficiency and total data throughput \cite{Wang2012,Willner2015,Milione2015,Ren2015}. One of the outstanding challenges to the practical application of spatial encoding is the slow rate at which distinct modes may be generated and detected. On the signal detection side, performing rapid and scalable spatial mode analysis is particularly desirable in the context of noisy and turbulent channels \cite{Lavery2017,Cox2019}, since fast and complete channel characterization is necessary for adaptive-optic noise compensation in real time \cite{Ren2014,Zhao2020a}.

Several approaches to generating spatial modes have been previously demonstrated, but none attain an ideal combination of speed, efficiency, scalability, and reconfigurability. Actively and arbitrarily switchable spatial mode generation is currently performed with adaptive optics such as spatial light modulators (SLMs) or digital micromirror devices (DMDs) which alter the phase and amplitude front of the beam \cite{Forbes2016}. Both of these have refresh rates on the order of $1 \un{kHz}$, which limits potential data transmission rates. This limitation is especially problematic for quantum key distribution (QKD) \cite{Sit2017} where the mode must be updated before every transmitted photon to ensure communication security. An alternative approach is to use multiple electro-optical modulators (EOMs) to illuminate one of several static phase plates, and recombine the resulting spatial modes using several beamsplitters \cite{Wang2012}. This method has the advantage of being very rapid (on the order of tens of $\mathrm{GHz}$), but the efficiency of such a device scales as $1/N$, where $N$ is the number of desired modes. The mode conversion efficiency can be increased from $\sim 1/N$  to a constant loss by the use of multi-plane light conversion (MPLC) \cite{Fontaine2019}, where the desired spatial mode is generated through several reflections from a suitably patterned SLM. In order to produce different spatial modes at different times, MPLC would still require a number of switches that scales as $N$. When $N$ becomes large, these approaches can become prohibitively expensive, limiting their scalability.

An acousto-optical modulator (AOM) is a common device that can act as a rapid, high-efficiency optical deflector with good multi-mode performance. AOMs are often used to control the propagation direction of light, similarly to how one would use a steerable mirror, acting like an N-port switch and allowing much greater efficiency and scaling capabilities than an $N$-port beamsplitter array \cite{Nikulin2008}. A double-pass AOM can be used to scan the frequency of a laser without affecting its spatial mode \cite{Donley2005}. In this configuration, different RF frequencies applied to the AOM cause the output beam to be steered and re-focused on different regions of a retroreflection mirror, which can be seen as a folded 4-f optical system. AOMs are also used to synthesize arrays of rapidly movable focal spots for trapping atoms \cite{Ksouri2014}. However, most often the spatial structure of the beam remains Gaussian, just like the input beam, with dynamical control exerted only over the beam's position or propagation direction.

Replacing the mirror of a folded 4-f system with a hologram (as shown in Figure \ref{fig:SpatialModeGenerationWithAOM}) allows different spatial modes to be encoded onto the reflected light when the input beam is steered to different regions of the hologram. By imprinting a linear phase grating onto the hologram, we can separate the light with the imprinted spatial mode structure from the zeroth order reflection. Thus, we can generate nearly-pure spatial modes of light with a rate limited only by the modulation rate of the AOM, which can readily reach $f = 1 \un{MHz}$. When realizing the hologram using an SLM with a resolution on the order of $10^{3}$ pixels, at least 10 modal patterns can be arranged side by side to select from $N=10$ distinct (and arbitrary) modes.

A similar approach to the present work has been taken by Radwell et al., demonstrating high-speed switching between arbitrary spatial patterns of light \cite{Radwell2014}. However, this earlier work only demonstrated mode generation at a speed of $10\un{kHz}$, which is potentially attainable with fast DMDs. In contrast, we observe switching rates for mode generation of up to $500 \un{kHz}$, well in excess of any existing SLM or DMD technology. Furthermore, while the modes generated in \cite{Radwell2014} were only characterized qualitatively through intensity measurements, we verify the phase structure of our generated modes quantitatively using a second hologram. In addition, we show that operating the setup in reverse allows for fast mode measurement, instead of fast mode generation. We demonstrate the projection of an unknown input state onto 3 pairs of mutually unbiased bases in a 2-dimensional Hilbert space, and use this capability to perform quantum state tomography of an unknown mode of light with a measurement rate in excess of $1\un{kHz}$ and an average reconstruction fidelity of $96.9\%$. 

\section{Experimental setup}

\begin{figure}[btp]
	\centering
	\includegraphics[width=1\linewidth]{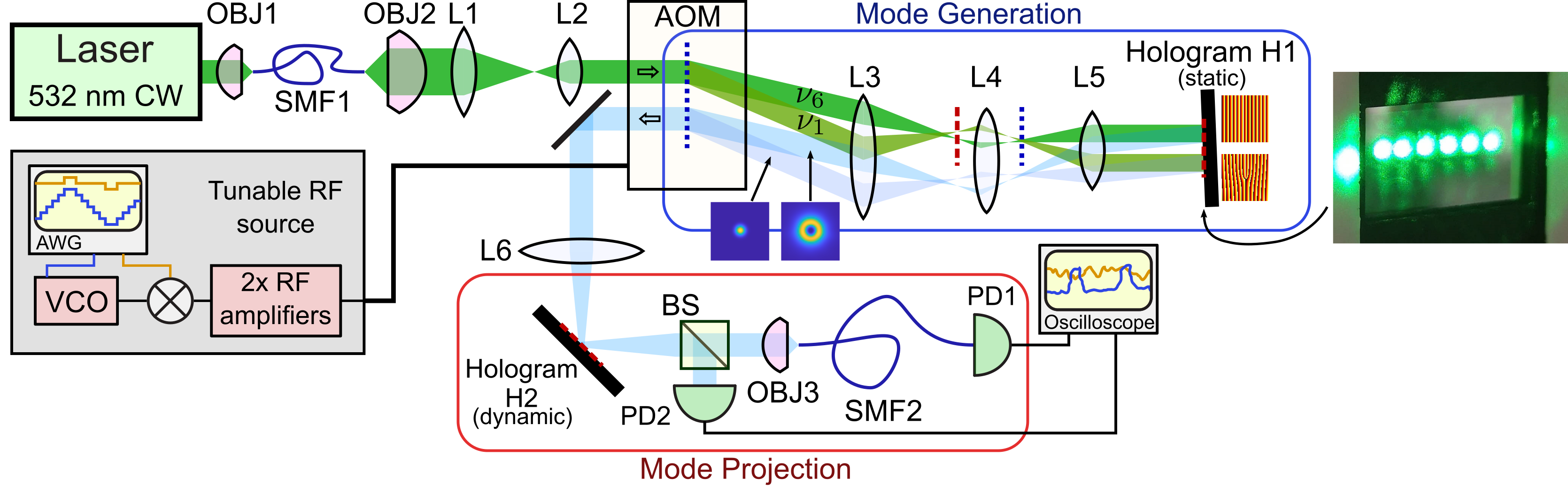}
	\caption{Scheme for rapidly switching between different spatial modes of light by using an AOM as a rapidly steerable mirror to multiplex multiple regions of a static hologram (H1). The generated modes are then analyzed using mode projection with a dynamic hologram (H2) and an SMF. Both holograms are implemented using SLMs. The blue dotted lines show the positions of object planes (near-field) of the AOM, while red dotted lines indicate the positions of the image planes (far field) of the AOM (which are also the object planes for the SLMs). Abbreviations used in the figure: acousto-optic modulator (AOM); beam sampler (BS); objective lens (OBJ); photodiode (PD); single-mode fiber (SMF); voltage-controlled oscillator (VCO).}
	\label{fig:SpatialModeGenerationWithAOM}
\end{figure}

The experimental setup is shown schematically in Figure \ref{fig:SpatialModeGenerationWithAOM}. Light from a $532\un{nm}$ CW laser is first coupled using an aspheric lens OBJ1 ($f = 4.51 \un{mm}$) into a single-mode fiber to spatially filter the laser output and ensure consistent alignment. The light is coupled back into free space using the objective OBJ2 ($f = 20 \un{mm}$), and is then sent through a telescope composed of lenses L1 and L2 ($f_1 = 200\un{mm}$, $f_2 = 50\un{mm}$). This telescope system ensures that the $1/e^2$ beam diameter of $0.8 \un{mm}$ is small enough to fit within the $2\un{mm}$ active aperture of the AOM, but is also large enough to produce a small spot size in the AOM far-field, where hologram H1 is located. The light then enters the AOM (Isomet 1205C-2) driven by RF signals with frequencies equally spaced between $\nu_1 = 120\un{MHz}$ and $\nu_6 = 70 \un{MHz}$. The RF signals are generated by an arbitrary waveform generator (AWG) that modulates a voltage-controlled oscillator (VCO), with an RF mixer used as a voltage-tunable attenuator to control the RF amplitude. The RF signal is amplified by two RF amplifiers before being sent to the AOM. The first-order diffraction from the AOM is deflected by approximately $0.15\un{mrad/MHz}$ at $532\un{nm}$, corresponding to a deflection angle between $10\un{mrad}$ and $18\un{mrad}$ at the RF frequencies used here. 


The optical system following the AOM (labeled ``Mode Generation'' in Fig. 1) is identical to a double pass AOM \cite{Donley2005} used for laser frequency modulation, except for H1 taking the place of the retro-reflection mirror. After leaving the AOM, the light enters another telescope, composed of lenses L3 and L4 (with focal lengths $f_3 = 250\un{mm}$ and $f_4 = 50\un{mm}$). This telescope acts to magnify the angular steering induced by the AOM five-fold, resulting in the full coverage of the width of H1. The light then passes through lens L5 ($f_5 = 300\un{mm}$), which converts the different diffraction angles from the AOM to different positions on H1. Next, the beam hits a part of the static hologram H1 corresponding to a particular desired output mode. We implement H1 using an SLM (Holoeye Pluto-2, resolution $1920\times1080$). The photograph inset to Fig. \ref{fig:SpatialModeGenerationWithAOM} shows the appearance of H1 with the incident light impinging on it, with the six spots corresponding to the six different RF frequencies applied to the AOM. Even though all six spots appear to be present simultaneously, this is due to the exposure time of the camera being longer than the sub-ms time required for the RF frequency to switch among the six settings $\nu_1\ldots \nu_6$. In fact, only one region of H1 is illuminated at any given time. We generate the desired spatial modes in the first diffracted order from the hologram on H1, with typical diffraction efficiency of $28\%$.

H1 is positioned such that the beam is not exactly normal to the plane of its SLM; rather, as illustrated in Fig. \ref{fig:SpatialModeGenerationWithAOM}, H1 is slightly tilted horizontally, causing the mode returning through the AOM to be slightly displaced from the incoming beam. The returning beam is then picked off with a D-shaped mirror and sent to the ``Mode Projection'' stage. The light passes through another lens L6 ($f_6 = 250\un{mm}$), which focuses it onto a second SLM (Cambridge Correlators SDE1024, resolution $1024\times768$), which displays the hologram H2 corresponding to the spatial mode onto which we are projecting the input light. The light then passes through a beam sampler (Thorlabs BSF10-A) at near-normal incidence, coupling about $3.9\%$ of the light onto a free space photodiode PD1 (Thorlabs PDA36A2), while the rest of the light is coupled to a second single mode fiber with objective OBJ3 ($f = 20\un{mm}$), which leads to an identical but fiber coupled photodiode PD2 (Thorlabs PDA36A2).

An arbitrary spatial mode is defined by its spatial profile, given by $A(x,y) e^{i\phi(x,y)}$, where $A(x,y)$ is the mode amplitude and $\phi(x,y)$ is the mode phase. A hologram for generating such a mode using a phase-only SLM, denoted as $\Phi_{\mathrm{SLM}}(x,y)$, can be found by superimposing the desired phase $\phi(x - x_0, y - y_0)$ with a linear phase grating, $k_x x + k_y y$, which separates the beam into both a first order diffraction and a zeroth-order diffraction \cite{Forbes2016}. Modulating the phase-only grating with the mode amplitude $A$ ensures that the spatial structure of the first-order diffracted beam closely approximates the desired spatial mode. Although more sophisticated approaches exist \cite{Bolduc2013}, a simple choice is to use the following SLM phase pattern:
\begin{equation}\label{eq:SLMPhasePattern}
    \Phi_{\mathrm{SLM}}(x,y) = A(x - x_0, y - y_0) \times \left(\phi(x - x_0, y - y_0) + k_x x + k_y y \mod 2\pi\right).
\end{equation}
The inverse grating periods $(k_x,k_y)$ and centre positions $(x_0,y_0)$ of the individual phase patterns on H1 were carefully optimized to correct for aberrations in the hologram and lenses L3, L4 and L5, to ensure that the light reaching H2 is precisely co-located for all generated modes. This alignment erases the information of which portion of H1 was used to produce any given spatial mode. The phase pattern given in \eqref{eq:SLMPhasePattern} was also used to perform mode projection with H2, although in that case, the inverse grating period $(k_x,k_y)$ and centre position $(x_0,y_0)$ were equal for all modes.

\begin{figure}[hbtp]
    \centering
    \setlength{\unitlength}{\columnwidth}
    \begin{picture}(1,0.27)
    \put(0.05,0.00){\includegraphics[height=0.25\columnwidth]{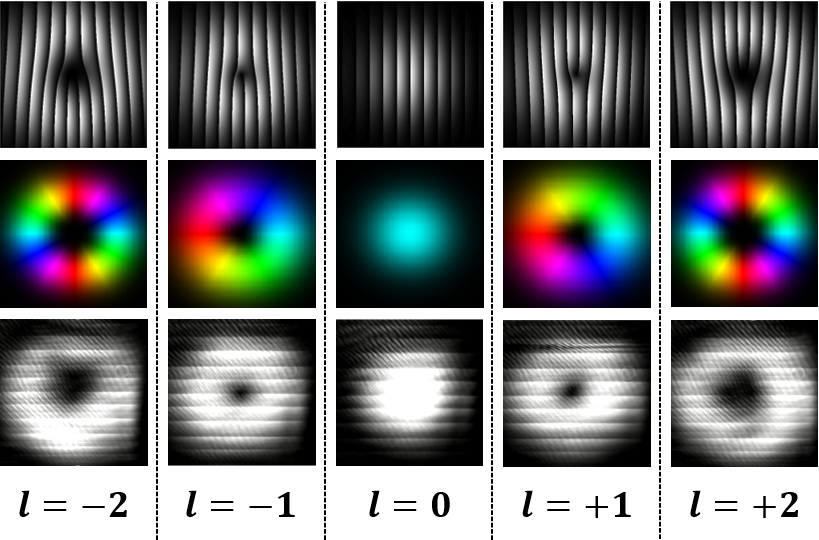}}
    \put(0.52,0.00){\includegraphics[height=0.25\columnwidth]{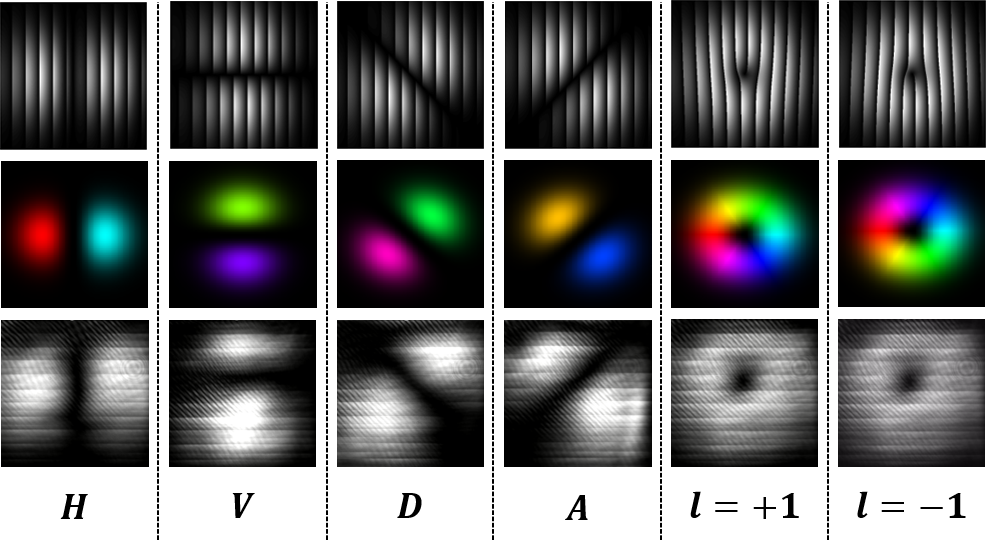}}
    \put(0.00,0.23){(a)}
    \put(0.47,0.23){(b)}
    \end{picture}
    \caption{Static phase patterns comprising hologram H1 are shown in the top row of images, with the middle row displaying the simulated mode profiles in the image plane (with hue indicating the phase and brightness the intensity of the field), and the bottom panel displaying the experimentally observed mode profiles generated by the AOM and H1. (a) Five OAM modes ($l_1= -2,-1,0,1,2$). (b) Six modes comprising 3 mutually unbiased bases in the 2-D Hilbert space spanned by the $l_1=\pm1$ states.}
    \label{fig:ModeGenSimulationWithImageLG_HG}
\end{figure}

\section{Rapid mode generation}

Figure \ref{fig:ModeGenSimulationWithImageLG_HG}(a) displays the static phase patterns comprising H1 used to generate 5 orthogonal orbital angular momentum (OAM) modes (Laguerre-Gauss beams with radial index $p=0$ and azimuthal index $l_1= -2,-1,0,1,2$), along with the mode profiles generated by this phase pattern in the far-field of H1. All of the grating patterns are displayed simultaneously in different regions of H1, with the AOM system alone being used to select the generated spatial mode. Figure \ref{fig:ModeGenSimulationWithImageLG_HG}(b) similarly shows the static phase patterns of H1 used to generate 6 states that form 3 mutually unbiased bases in the 2-D Hilbert space spanned by the $l_1=\pm1$ states. These 6 states consist of the two $l_1=\pm1$ OAM modes, along with horizontal, vertical, diagonal and anti-diagonal variations of the TE10 Hermite-Gauss mode. Figure \ref{fig:ModeGenSimulationWithImageLG_HG}(b) shows the simulated and imaged mode profiles generated by these phase patterns. The observed mode profiles are well aligned, with the same center position and propagation direction, which is necessary for the modes shown in Fig. \ref{fig:ModeGenSimulationWithImageLG_HG}(a) to be orthogonal and for the modes in Fig. \ref{fig:ModeGenSimulationWithImageLG_HG}(b) to form 3 mutually unbiased bases.

\begin{figure}[hbtp]
    \centering
    
    \setlength{\unitlength}{\columnwidth}
    \begin{picture}(1,0.38)
    \put(0.00,0.00){\includegraphics[width=0.5\columnwidth]{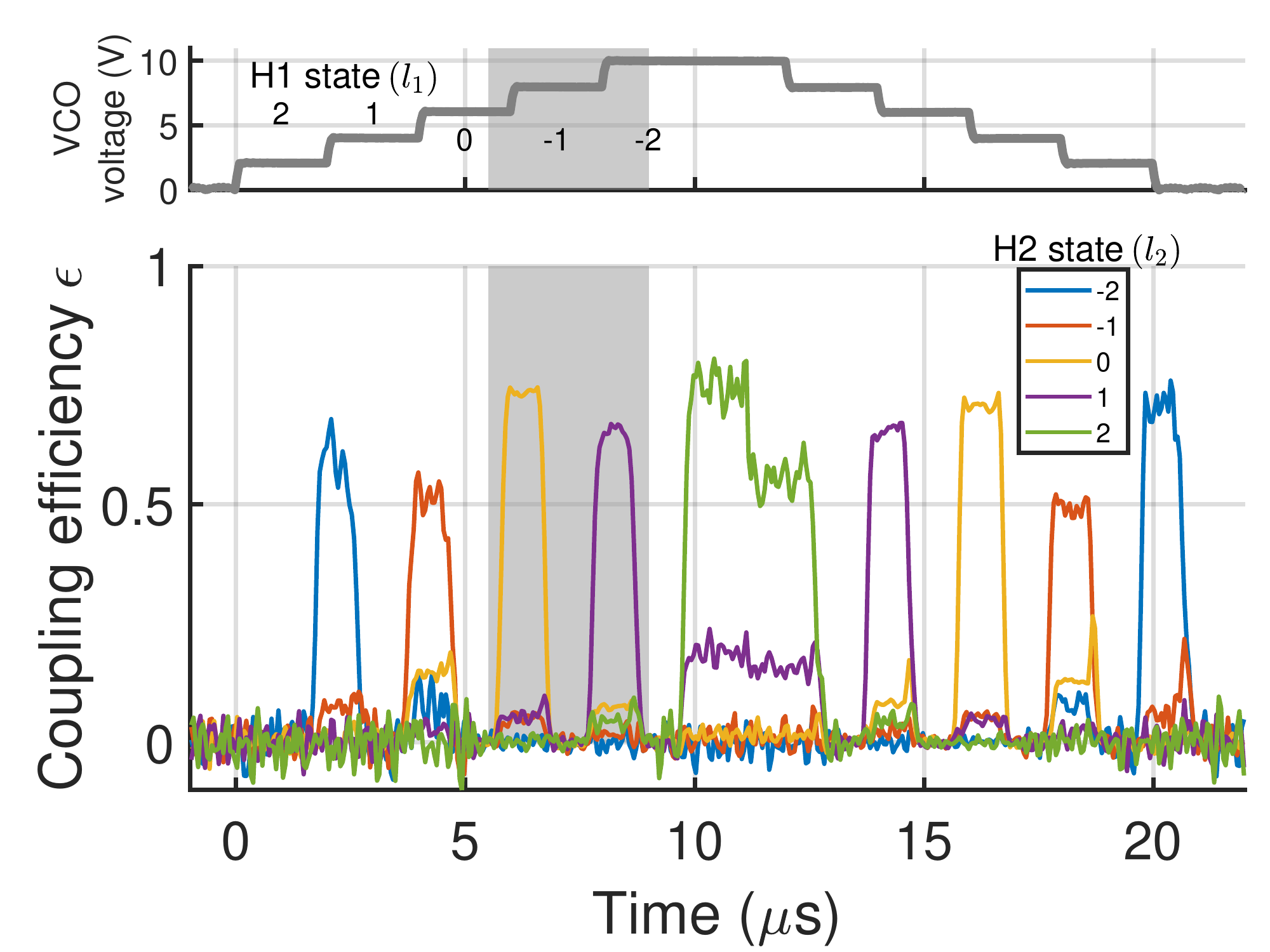}}
    \put(0.50,0.00){\includegraphics[width=0.5\columnwidth]{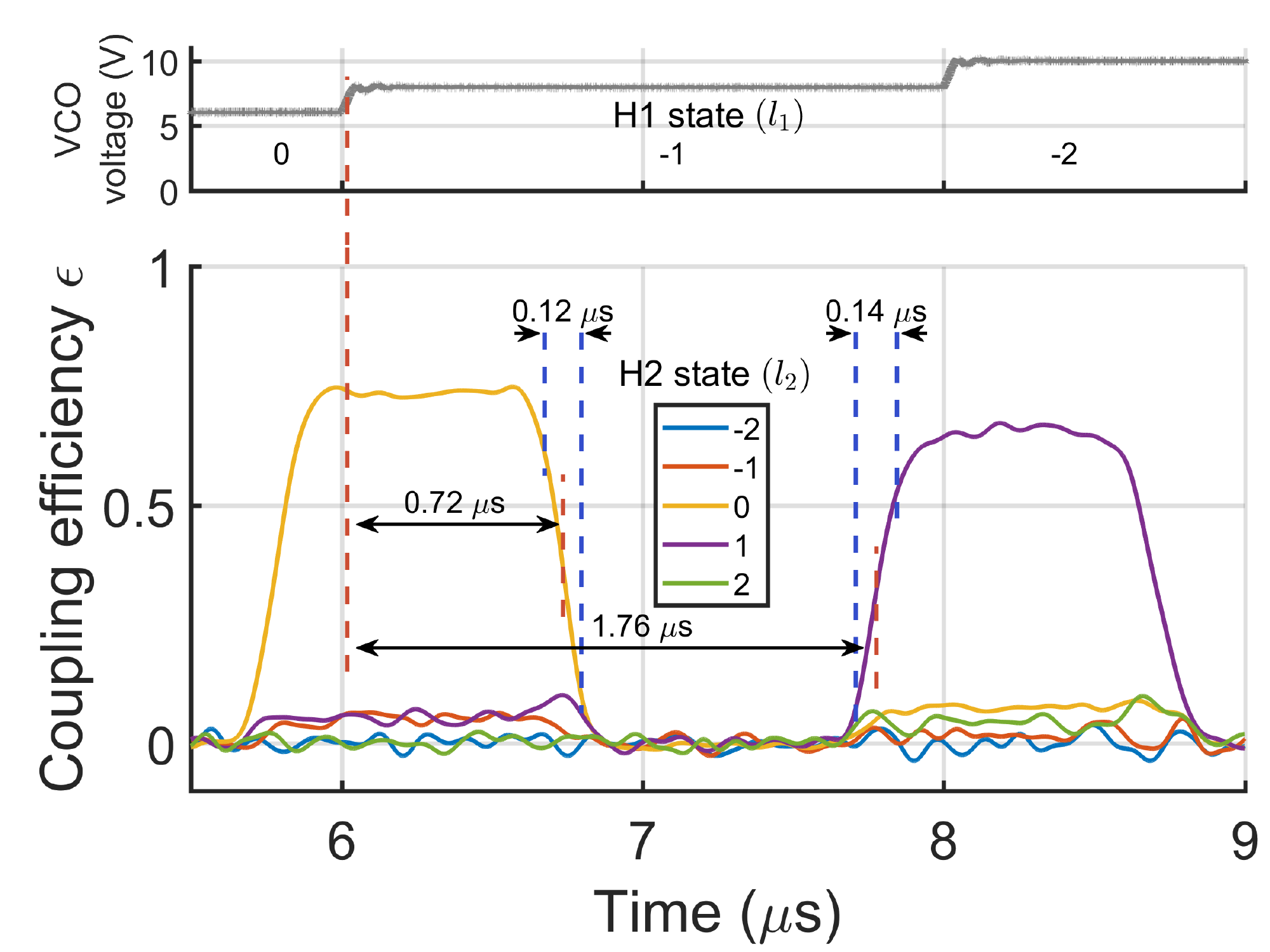}}
    \put(0.00,0.36){(a)}
    \put(0.50,0.36){(b)}
    \end{picture}
    \caption{Single-shot measurements of mode projection coupling efficiency for modes generated in the OAM mode basis at a rate of $500 \un{kHz}$, with the corresponding VCO voltage graphs displayed above. A trace covering two full sweeps over the 5 generated modes is displayed in (a), with the gray highlighted region shown in more detail in (b).}
    \label{fig:ModeProjection500k}
\end{figure}

We characterize the quality and switching speed of mode generation by performing mode projection using H2. We fix the pattern on H2 to one that would convert a Gaussian input beam into an output beam with OAM value $l_2$. Consequently, only input modes with $l_1=-l_2$ are converted into a Gaussian beam that can couple efficiently into the fiber (SMF2). We find the coupling efficiency $\epsilon$ from the photodiode voltages $V_1$ and $V_2$ and the beam sampler splitting ratio $\eta = 3.9\%$:
\begin{equation}\label{eq:CouplingEfficiencyDefinition}
\epsilon = \frac{1-\eta}{\eta} \frac{V_1}{V_2}
\end{equation}
The generation of a high-fidelity mode with OAM $l_1$ is confirmed by a value of $\epsilon$ close to unity when $l_1=-l_2$, and near-zero otherwise.

We step the RF frequency between $5$ values, ranging from $\nu_1 = 120 \un{MHz}$ and $\nu_5 = 80 \un{MHz}$, at a switching rate of up to $500 \un{kHz}$, generating each of the $l_1=-2,-1,0,1,2$ OAM modes in order. Figure \ref{fig:ModeProjection500k}(a) shows the measured traces of the coupling efficiency $\epsilon$ into each of the corresponding output modes. As expected, $\epsilon$ is high only when $l_1=-l_2$, indicating that the generated modes closely match the desired OAM states. The noise in $\epsilon$ is greater for times below $5 \un{\mu s}$, between $10$ and $13 \un{\mu s}$, and above $19 \un{\mu s}$ due to the lower diffraction efficiency by the AOM and H1 for those spatial modes, which results in a lower voltage $V_2$ on the free-space photodiode that acts to magnify any noise or fluctuation in the fiber-coupled photodiode voltage $V_1$. Figure \ref{fig:ModeProjection500k}(b) shows a more detailed version of the gray highlighted part in Fig. \ref{fig:ModeProjection500k}(a). There is a $0.72 \un{\mu s}$ delay between when the AWG ramps up the VCO voltage to output the RF frequency for $l_1=-1$, and the time the previously generated $l_1=0$ mode is switched off. Following an additional delay of $1.04\un{\mu s}$, the desired $l_1=-1$ mode switches on, giving a total delay of $1.76\un{\mu s}$ between the moment the AWG selects mode $l_1=0$ and the moment this mode is first produced by the AOM and H1. 

The delays we observe are explained by the finite speed of sound through the AOM crystal, $v_s = 3.63 \un{km/s}$. At this speed, it takes sound waves $0.72 \un{\mu s}$ to cover the $2.6\un{mm}$ distance between the  piezoelectric transducer in the AOM and the first pass of the light through the AOM. The additional delay of $1.04\un{\mu s}$ is due to the $3.7 \un{mm}$ separation between the two passes of the light through the AOM. While these delays limit the modulation frequency in this particular experimental configuration to around $1\un{MHz}$, they are not fundamental. By aligning both passes of the light through the AOM to be within $0.5\un{mm}$ of the transducer, it would be possible to reduce all delays in the system down to $140 \un{ns}$, permitting modulation rates of up to $5\un{MHz}$. 

\begin{figure}[hbtp]
    \centering
    \setlength{\unitlength}{\columnwidth}
    \begin{picture}(1,0.38)
    \put(0.07,0.01){\includegraphics[width=0.42\columnwidth]{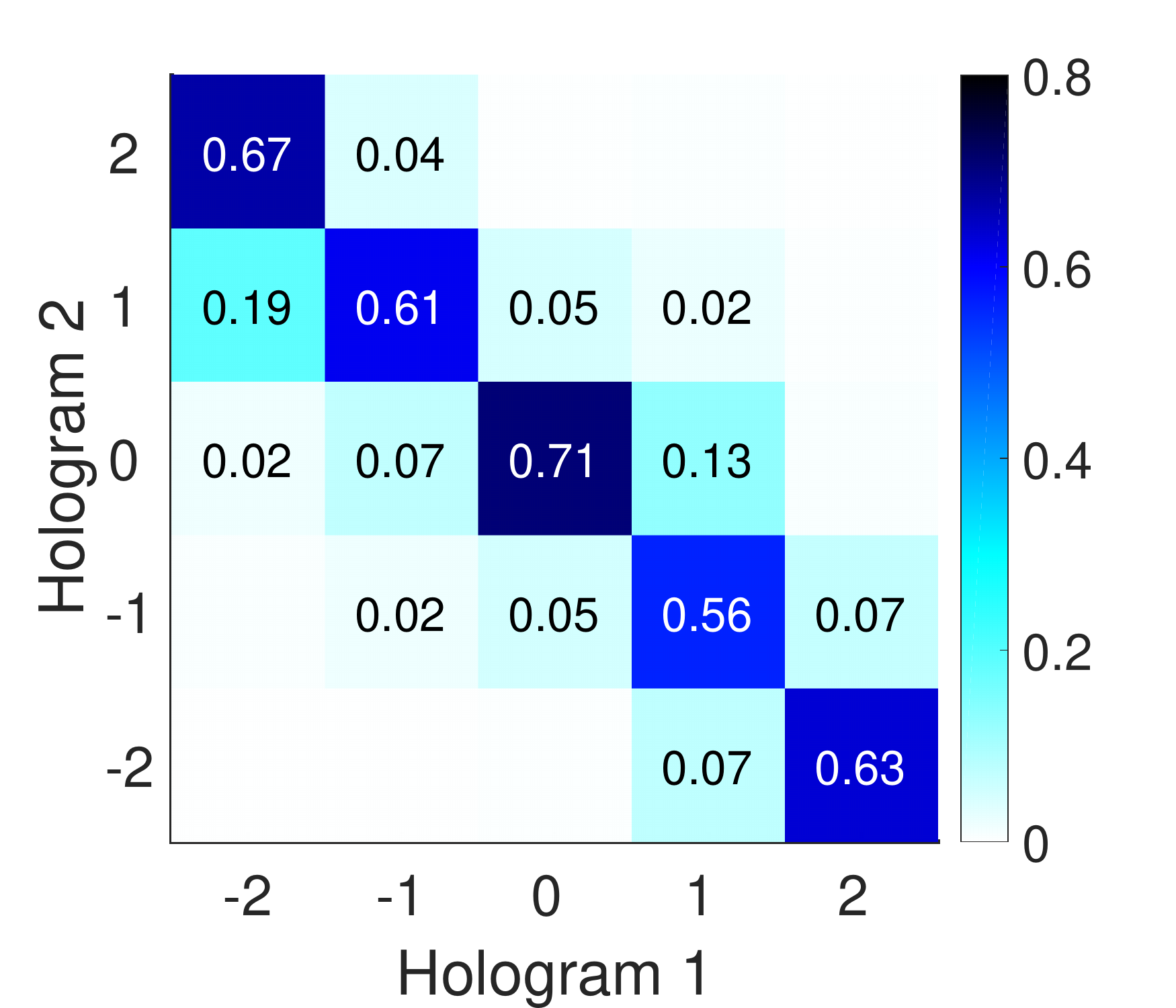}}
    \put(0.5,0.00){\includegraphics[width=0.5\columnwidth]{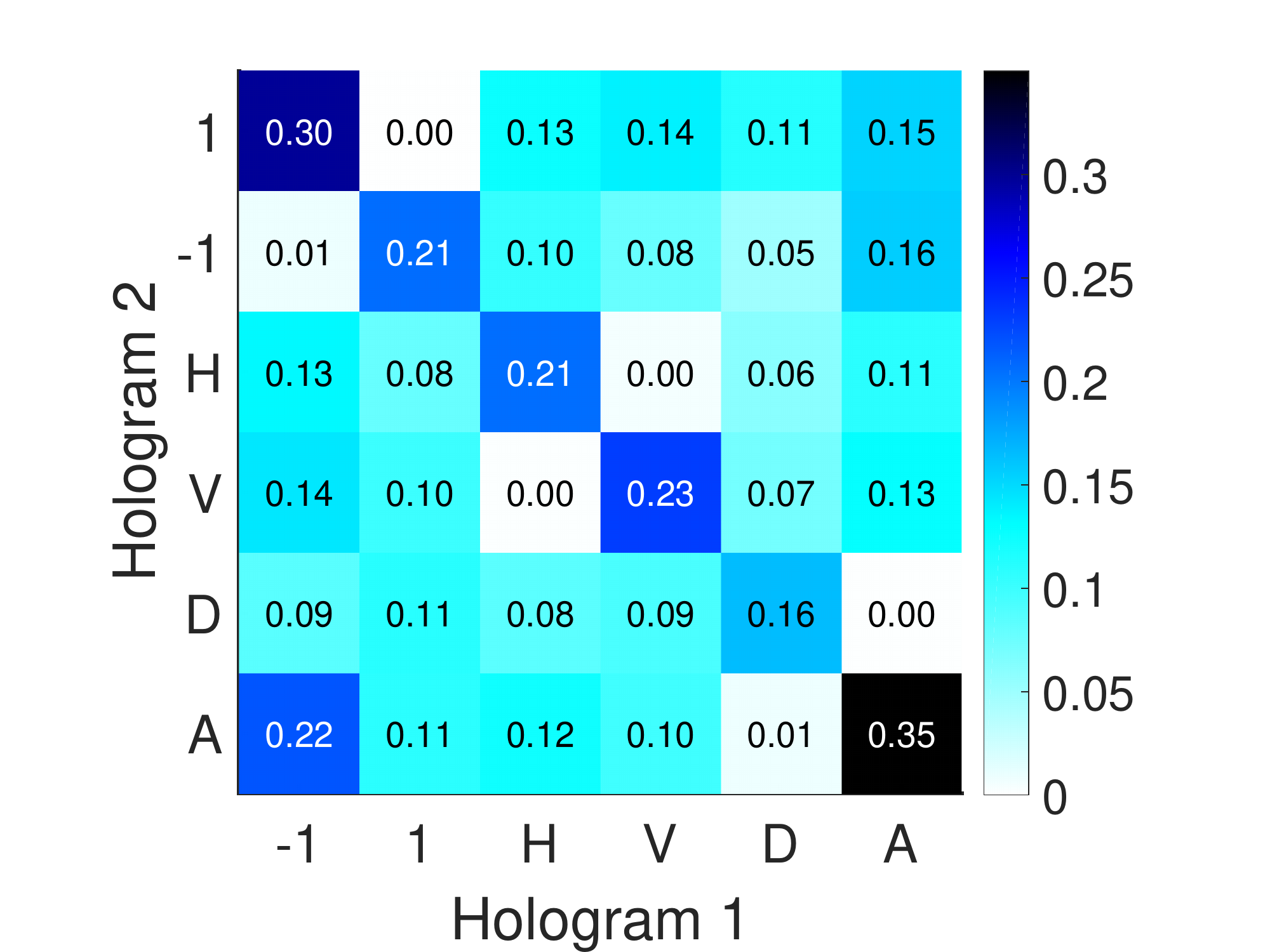}}
    \put(0.07,0.35){(a)}
    \put(0.53,0.35){(b)}
    \end{picture}
    \caption{Cross-talk matrices showing the overlap $\left|\braket{\psi_2|\psi_1}\right|^2$ between modes generated with the AOM and H1 and ones projected using H2. (a) OAM states shown in Fig. \ref{fig:ModeGenSimulationWithImageLG_HG}(a). (b) Six states comprising 3 mutually unbiased bases shown in Fig. \ref{fig:ModeGenSimulationWithImageLG_HG}(b).
    }
    \label{fig:CrossTalkMatrices}
\end{figure}

The coupling efficiency traces shown in Fig. \ref{fig:ModeProjection500k}(a) contain information about the mode overlap between every pair of OAMs $l_1$ and $l_2$, which can be expressed as the cross-talk matrix shown in Fig. \ref{fig:CrossTalkMatrices}(a). The average overlap of the generated modes with the target modes is equal to $63.6\%$, while the average cross-talk to each of the other modes equals $3.7\%$. The mode generation fidelity is limited by imperfections in the two SLMs, especially the nonlinear dependence of the optical phase shift on the gray value of the computer-generated holograms shown in the top row of Fig. \ref{fig:ModeGenSimulationWithImageLG_HG}(a). The fidelity could be improved by calibration of the SLMs used to implement the holograms \cite{Xun2004}, or through the use of binary holograms \cite{Forbes2016}.

If the mode projection on H2 was also accelerated by an AOM, the cross-talk matrix in Fig. \ref{fig:CrossTalkMatrices}(a) could be measured at a rate of $20 \un{kHz}$. This rate is more than two orders of magnitude faster than the Greenwood frequency of around $60\un{Hz}$ \cite{Greenwood1977,Tyler1994}, the characteristic correlation time of turbulence-induced distortion through atmospheric channels. Therefore, our system is suitable for characterizing such channels when performing real-time adaptive optics correction, and could be used to further improve the transmission fidelity for both classical \cite{Ren2014} and quantum \cite{Zhao2020a} communication.

\section{Rapid quantum state tomography}

An unknown quantum state $\rho$ in a $2$-dimensional Hilbert space can reconstructed by measuring the overlap of $\rho$ with each of six states comprising three mutually unbiased bases, analogously to characterizing a polarization state by measuring its three Stokes parameters. In the case of the space spanned by the OAM states $l=1$ and $l=-1$, these six states are the ones shown in Fig. \ref{fig:ModeGenSimulationWithImageLG_HG}(b):
\begin{equation} \label{eq:SixMUBStates}
    \ket{+1}; \ket{-1}; \; \ket{H,V} = \frac{\ket{1}\pm\ket{-1} }{\sqrt{2}}; \; \ket{D,A} = \frac{\ket{1}\pm i \ket{-1} }{\sqrt{2}}.
\end{equation}
Thus, we can use the same experimental configuration as before to recover an unknown spatial mode corresponding to the hologram H2. We encode each of six states given in \eqref{eq:SixMUBStates} on different regions of H1, use the AOM to switch among them, and measure the resulting coupling efficiencies into SMF2: $\epsilon_{+1}$, $\epsilon_{-1}$, $\epsilon_{H}$, $\epsilon_{V}$, $\epsilon_{D}$, $\epsilon_{A}$. From these efficiencies, we determine the overlap of the unknown state $\rho$ with the Pauli operators $\sigma_i$:
\begin{eqnarray}
S_x &=& \mathrm{tr}(\sigma_x \rho) = \frac{\epsilon_{H} - \epsilon_{V}}{\epsilon_{H} + \epsilon_{V}} \nonumber\\
S_y &=& \mathrm{tr}(\sigma_y \rho) = \frac{\epsilon_{D} - \epsilon_{A}}{\epsilon_{D} + \epsilon_{A}} \\
S_z &=& \mathrm{tr}(\sigma_z \rho) = \frac{\epsilon_{+1} - \epsilon_{-1}}{\epsilon_{+1} + \epsilon_{-1}} \nonumber
\end{eqnarray}
which in turn implies $\rho = S_x \sigma_x + S_y \sigma_y + S_z \sigma_z$.
To confirm the validity of these idealized formulas, we first realize each of the six states from \eqref{eq:SixMUBStates} on H2, measuring the cross-talk matrix shown in Figure \ref{fig:CrossTalkMatrices}(b). We see that within each pair of complementary states (i.e. each $2\times 2$ square along the diagonal) the mode cross-talk is very low, and thus the corresponding spin projections $S_i$ are very near to their ideal values of $\pm 1$. 

\begin{figure}[hbtp]
    \centering
    \setlength{\unitlength}{\columnwidth}
    \begin{picture}(1,0.42)
    \put(0.00,0.00){\includegraphics[width=0.50\columnwidth]{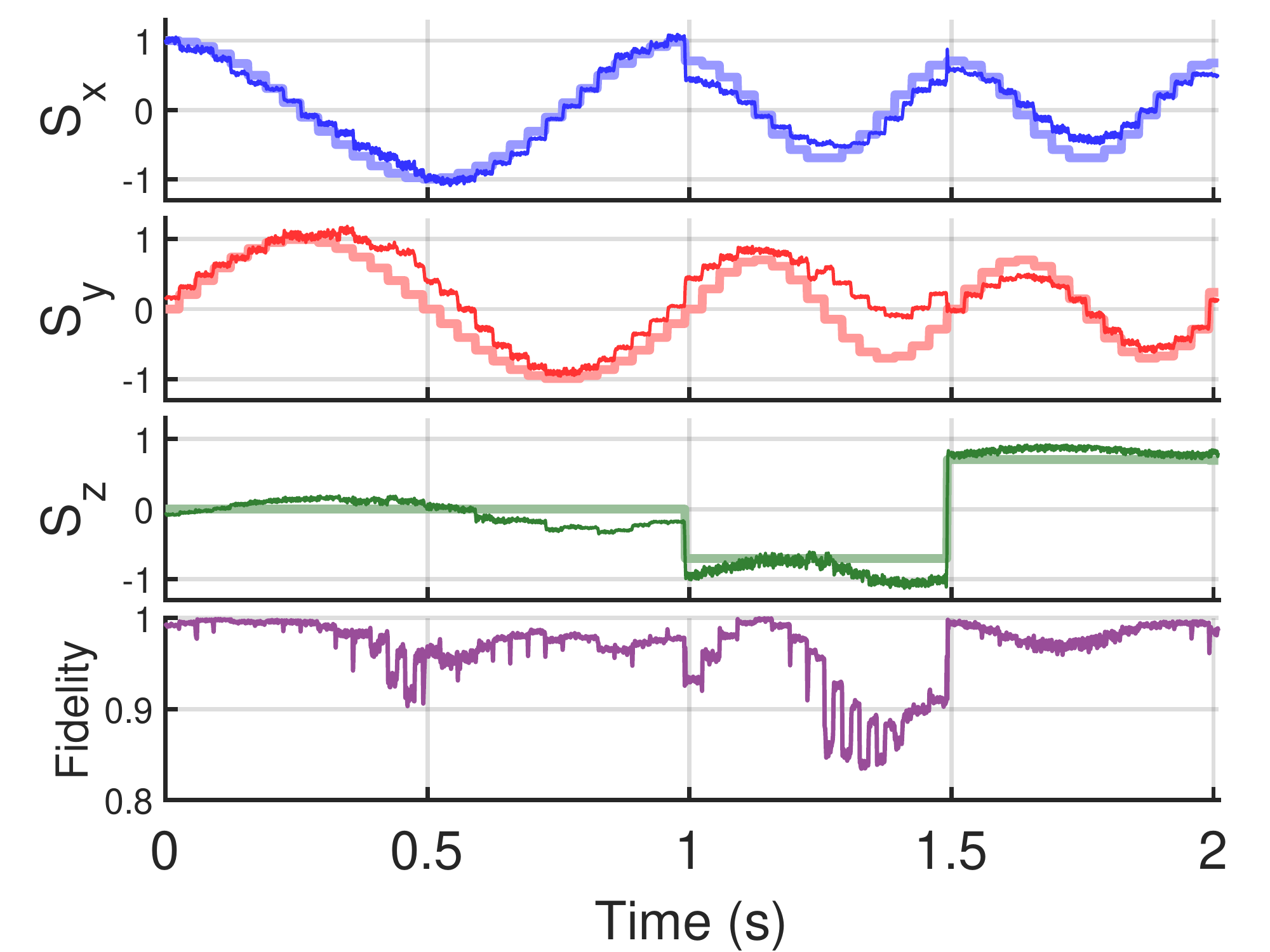}}
    \put(0.50,0.00){\includegraphics[width=0.50\columnwidth]{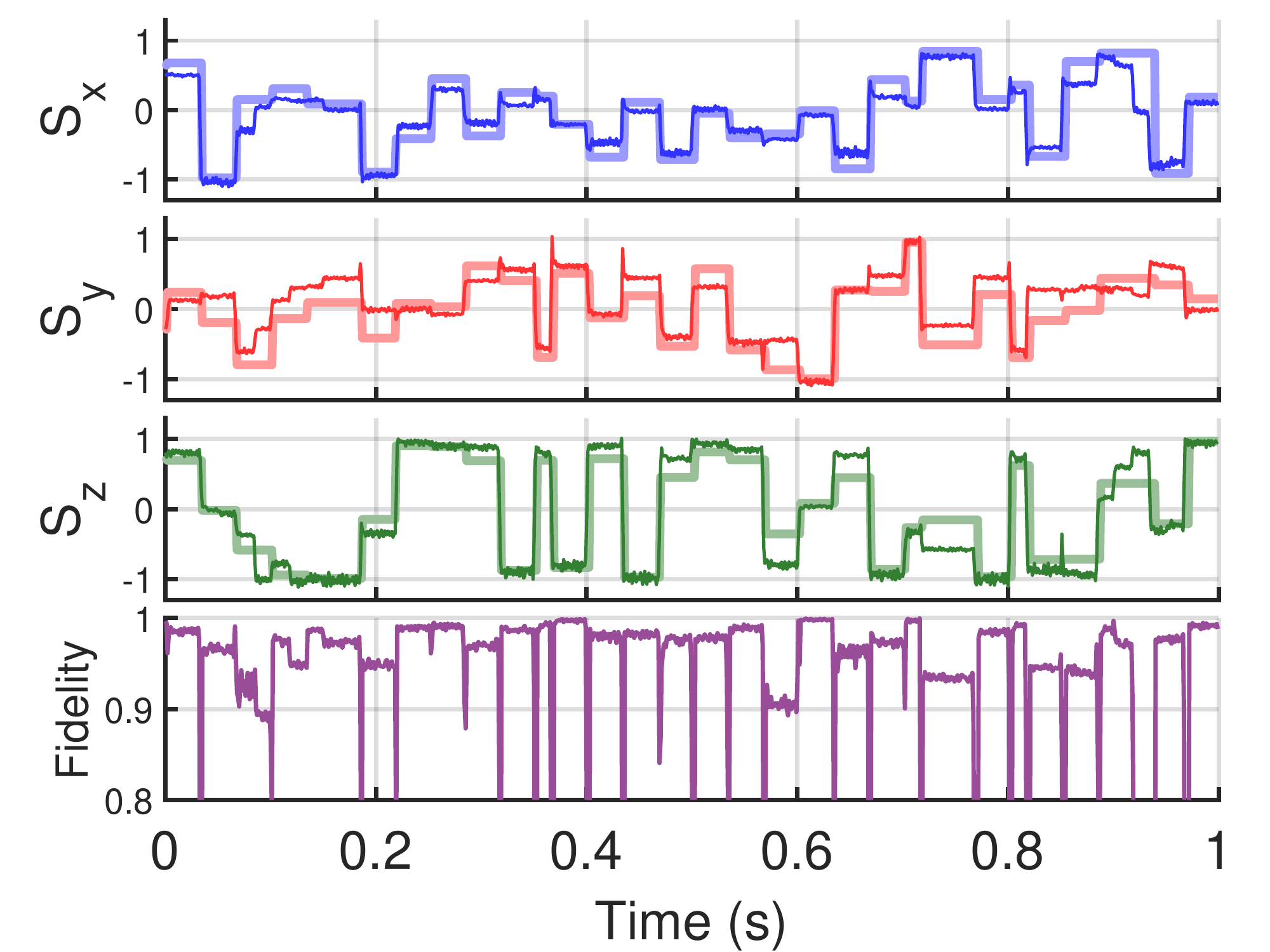}}
    \put(0.00,0.36){(a)}
    \put(0.50,0.36){(b)}
    \end{picture}
    \caption{Comparison of measured states (thin dark lines) against generated states (wide light lines) in a 2-D Hilbert space. The parameters $S_x$, $S_y$ and $S_z$ are the $x$, $y$ and $z$ coordinates of the state on the Bloch sphere, with the bottom panel displaying the fidelity between the generated and measured state. Panel (a) depicts a continuous trajectory wrapping around the Bloch sphere three times -- once around the equator ($S_z = 0$), and once each around the $\pm 45^\circ$ parallels ($S_z = \pm1/\sqrt{2}$), while (b) depicts a random trajectory.}
    \label{fig:QuantumStateTomography}
\end{figure}

To show that this approach is able to retrieve arbitrary modes with high fidelity, we generate a set of holograms H2 that are stored as a video file to ensure fast and consistent timing when displayed on the SLM. The first part of the video encodes a trajectory that encircles the Bloch sphere three times -- once around the equator ($S_x = \cos (\omega t)$, $S_y = \sin (\omega t)$, $S_z = 0$), and once each around the $\pm 45^\circ$ parallels ($S_x = \cos (\omega t)/\sqrt{2}$, $S_y = \sin (\omega t)/\sqrt{2}$, $S_z = \pm1/\sqrt{2}$). The comparison of the theoretically expected and experimentally measured states is shown in Fig. \ref{fig:QuantumStateTomography}(a), achieving an average fidelity of $96.8\%$. We then repeat the experiment with a sequence of 30 random states, as shown in Fig. \ref{fig:QuantumStateTomography}(b), where we also see a high average fidelity of $97.1\%$. It is important to note that all the measurements were performed in a single shot, without re-playing the patterns on H2.

\begin{figure}[hbtp]
    \centering
    \includegraphics[width=0.7\columnwidth]{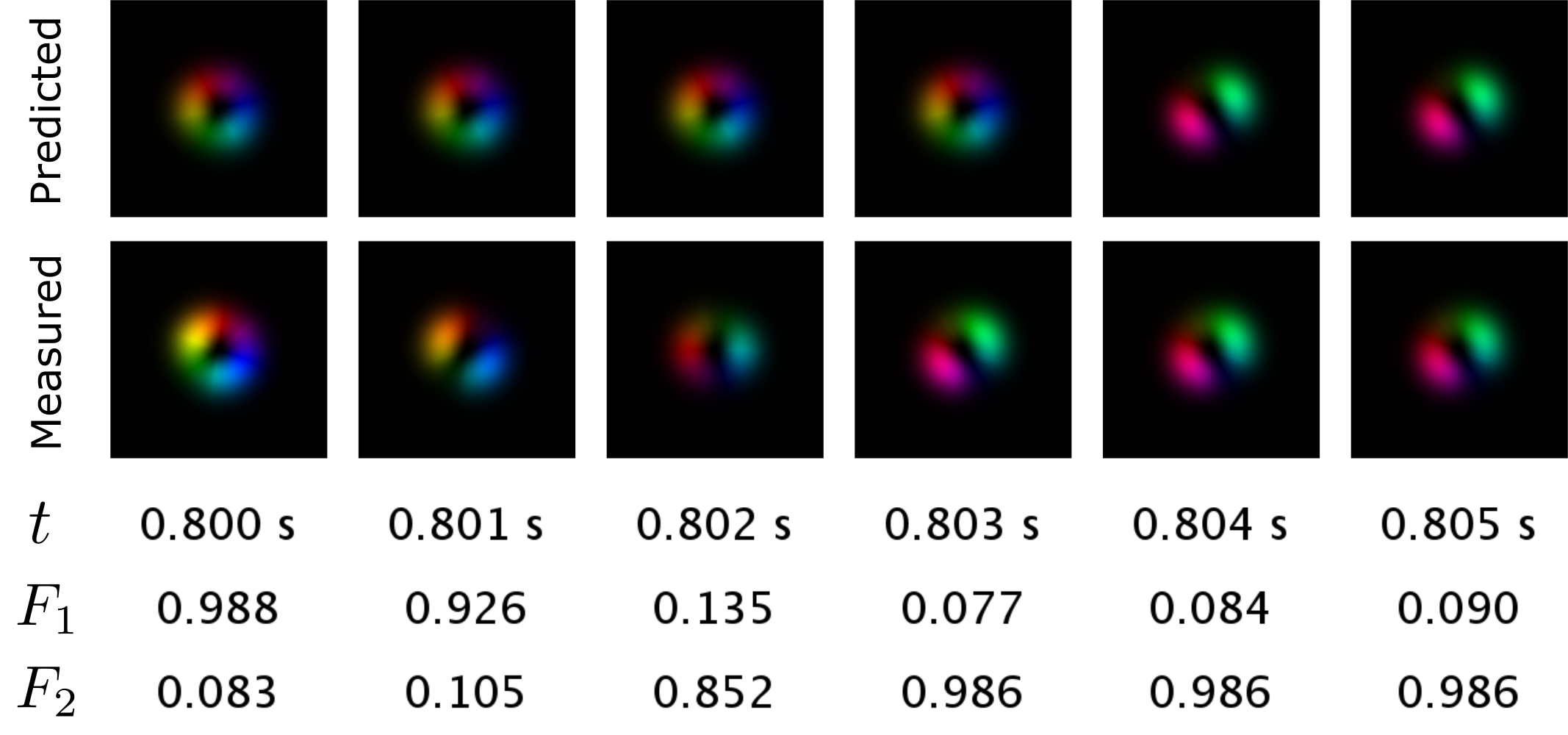}
    \caption{Comparison of the predicted spatial modes generated by H2, assuming instantaneous updating of the displayed image (top row of images) to the spatial modes measured with the AOM and H1 (bottom row of images). $F_1$ indicates the fidelity between the reconstructed mode to mode expected at the start ($t=0.800\un{s}$) while $F_2$ is the fidelity to the mode expected at the end ($t=0.805 \un{s}$). Modes and times correspond to the H2 trajectory shown in Fig. \ref{fig:QuantumStateTomography}(b).}
    \label{fig:ReconstructedVsActualModes}
\end{figure}

For these measurements, the AOM was switching between the six modes at a rate of only $10\un{kHz}$, due to memory limitations on the oscilloscope (otherwise, the full $500 \un{kHz}$ switching speed demonstrated earlier could have been utilized). This means we are able to retrieve a new estimate of the state on H2 at a rate of $1.33 \un{kHz}$, much faster than the display refresh rate of the SLM ($60 \un{Hz}$) and fast enough to capture the transition dynamics between the different modes being generated by H2. Figure \ref{fig:ReconstructedVsActualModes} illustrates one such transition near $t=0.800 \un{s}$ in Fig. \ref{fig:QuantumStateTomography}(b). We can see that even though most of the transition between the initial and final modes occurs in under $1\un{ms}$, it takes about $3\un{ms}$ for the transition to complete. We can also see the effect of timing fluctuations in the playback of the video on the SLM -- even though we expect the mode transition to occur around $t=0.803 \un{s}$, it is actually closer to $t=0.801 \un{s}$. Similar signatures can be seen in Fig. \ref{fig:QuantumStateTomography}(b). Even though we are playing back a video with a nominal frame rate of $30 \un{Hz}$ on an SLM with a nominal frame rate of $60 \un{Hz}$, and thus expect each hologram to be displayed for a duration of $33.3 \un{ms}$, we find that several frames are displayed for either $16.7 \un{ms}$ or $50.0 \un{ms}$. We also find that occasionally the SLM produces holograms interpolating between two adjacent video frames, likely a consequence of the graphics card smoothing out the video output to the SLM for H2. While the transient behaviors we observe can be neglected in experiments with slow SLM refresh rates, they can become significant for those experiments trying to make full use of the switching speed of SLMs, such as for high-speed data transmission, turbulence compensation \cite{Ren2014,Zhao2020a}, or dynamical atomic trapping \cite{Stuart2018}.

\section{Discussion}

Our measurements show that an AOM can be effectively used to multiplex different regions of a static hologram realized with an SLM, resulting in spatial mode switching rates far in excess of what can be achieved with the SLM alone. To ensure the high-fidelity generation of spatial modes, it is essential for the output modes after the second pass through the AOM to be co-located and co-propagating relative to one another, regardless of the RF frequency applied to the AOM.

That is, for each diffraction angle that the AOM applies to a beam upon its first pass, the beam must exit the AOM on its second pass in the same location, with the same transverse momentum. 
While a folded 4-$f$ system we implement closely approximates this idealized situation, it is impossible to eliminate all aberrations within the setup, particularly in the position and orientation of H1 and lenses L3, L4 and L5. We were able to compensate for these aberrations by optimizing the precise position of each pattern comprising H1, as well as the period and angle of the diffraction grating for each mode.

Here, we have demonstrated the switching of $6$ modes with a rate of up to $500 \un{kHz}$. The switching speed could be further improved by aligning both passes of the light through the AOM closer to the transducer, which would allow rates in excess of $5 \un{MHz}$. The number of multiplexed modes could be increased by using a higher-resolution SLM, or by using two AOMs (or a 2-D AOM) to produce a grid of spots on the SLM, rather than a single row. By using both of these techniques, we estimate that up to $100$ modes could be generated with a single SLM, and even thousands of modes could potentially be multiplexed by using a static hologram produced through nanofabrication. Using a nanofabricated hologram rather than an SLM would have the additional benefit of greatly increasing the damage threshold of the device, allowing its use for high-power and pulsed lasers.

This method of AOM-based hologram multiplexing could be further extended and applied in several ways. The hologram H1 could be re-written while the AOM scans the beam across the corresponding SLM, allowing for producing a high-rate stream of arbitrary modes (rather than ones picked out of a fixed library). For example, if the SLM can be updated at a rate of $100 \un{Hz}$, and encode 100 different spatial modes in its different regions, the mode generation rate can be now effectively increased up to $10 \un{kHz}$.
Free-space turbulence could be characterized at a very high rate by using two such devices, one on the transmitter side, and the other on the receiver side, to perform quantum process tomography of the channel. A full-rank quantum process tomography for a $d=10$ dimensional channel (describing arbitrary transformations of density matrices), which requires $d^4$ separate measurements \cite{Mohseni2008}, could be performed in under $10 \un{ms}$ with this approach.
Instead of multiplexing spatial patterns on a single SLM, the AOM could also be used as the front-end to a MPLC device that converts an array of Gaussian beams into a chosen set of spatial modes, for example Laguerre-Gauss modes \cite{Fontaine2019}. This combination of an AOM and MPLC would allow for high-fidelity, high-speed, and high-efficiency generation of any of the spatial modes the MPLC is capable of producing, coming close to achieving the ideal combination of speed, efficiency, scalability, and reconfigurability for devices controlling spatial modes of light.

\section{Conclusion}

By using an AOM in a double-pass configuration to multiplex different regions of a static hologram implemented using an SLM, we have demonstrated both the rapid generation and sorting of spatial modes of light. We switch up to $6$ modes with a rate of up to $500 \un{kHz}$, and apply this device to perform quantum state tomography in a 2-dimensional Hilbert space at a rate exceeding $1\un{kHz}$ and an average fidelity of 96.9\%. This state tomography is fast enough to directly observe the dynamics of the displayed pattern updating on another SLM.
While our implementation focused on a small number of OAM spatial modes, in principle arbitrary spatial mode bases could be generated. With straightforward technical improvements, switching between $100$s of modes at a rate above $1 \un{MHz}$ could be attained.

\section*{Acknowledgements}

This research was funded by National Science and Engineering Research Council of Canada (award RGPIN/2017-06880) and the US Office of Naval Research (award N00014-17-1-2443). B.B. acknowledges the support of the Banting Postdoctoral Fellowship. R.W.B. acknowledges the support of the Canada Research Chair Program.

\section*{Disclosures}

The authors declare no conflicts of interest.


\bibliography{SpatialModesWithAOM_20200730}
\bibliographystyle{osajnl}

\end{document}